\documentclass[aps,prl,showpacs,preprintnumbers,amsmath,amssymb]{revtex4-2}
\usepackage{hyperref}
\usepackage[dvipdfmx]{graphicx}
\usepackage{amsmath}
\usepackage{graphicx,float,color}

\definecolor{c1}{RGB}{0,148,17}
\definecolor{c2}{RGB}{255,137,0}

\newcommand{\be}{\begin{equation}}
\newcommand{\ba}{\begin{eqnarray}}
\newcommand{\ea}{\end{eqnarray}}
\newcommand{\ee}{\end{equation}}

\newcommand{\bea}{\begin{eqnarray}}
\newcommand{\eea}{\end{eqnarray}}
\newcommand{\bes}{\begin{equation*}}
\newcommand{\beas}{\begin{eqnarray*}}
\newcommand{\eeas}{\end{eqnarray*}}
\newcommand{\bas}{\begin{array*}}
\newcommand{\eas}{\end{array*}}
\newcommand{\ees}{\end{equation*}}

\newcommand{\bpm}{\begin{pmatrix}}
\newcommand{\epm}{\end{pmatrix}}
\newcommand{\bbm}{\begin{bmatrix}}
\newcommand{\ebm}{\end{bmatrix}}

\usepackage{bm}

\begin{document}

\begin{flushleft}
IPM/P-2021/018 ; YITP-21-66 ; MPP-2021-100
\end{flushleft}

\title{On the Time Scaling of Entanglement in Integrable Scale-Invariant Theories}
\author{M. Reza Mohammadi Mozaffar$^{a,b}$ and Ali Mollabashi$^{c,d}$}

\affiliation{
$^a$ Department of Physics, University of Guilan, P.O. Box 41335-1914, Rasht, Iran
\\
$^b$ School of Physics, Institute for Research in Fundamental Sciences (IPM),
19538-33511, Tehran, Iran
\\
$^c$
Center for Gravitational Physics,
Yukawa Institute for Theoretical Physics, Kyoto University,
Kitashirakawa Oiwakecho, Sakyo-ku, Kyoto 606-8502, Japan
\\
$^d$
Max-Planck-Institut for Physics,
Werner-Heisenberg-Institut, 80805 Munich, Germany
}

\date{\today}

\begin{abstract}
In two dimensional isotropic scale-invariant theories, the time scaling of the entanglement entropy of a segment is fixed via the conformal symmetry. We consider scale invariance in a more general sense and show that in integrable theories that the scale invariance is anisotropic between time and space, parametrized by $z$, most of the entanglement is carried by the \textit{slow modes} for $z>1$. At early times entanglement grows linearly due to the contribution of the fast modes, before smoothly entering a slow mode regime where it grows forever with $t^{\frac{1}{1-z}}$. The slow mode regime admits a logarithmic enhancement in bosonic theories. We check our analytical results against numerical simulations in corresponding fermionic and bosonic lattice models finding extremely good agreement.
We show that in these non-relativistic theories that the slow modes are dominant, local quantum information is universally scrambled in a stronger way compared to their relativistic counterparts.  
\end{abstract}
\maketitle

\section{Introduction}
Understanding the dynamics of entanglement is a central problem on the interface of statistical physics, condensed matter physics, quantum field theory, quantum information, and gravitational physics. Among a wide set of theoretical questions tied with this interdisciplinary topic are thermalization and relaxation of many-body systems, dynamics of quantum phase transitions, and evaporation of black-holes (see e.g. \cite{Reviews}).
Besides these theoretical interests, in recent years the revolutionary experiments with cold atoms made it possible to experimentally probe different features of closed quantum systems (see e.g. \cite{Experimetal}).

In the quantum field theory context, the conformal symmetry is strong enough to fix the dynamics of entanglement for certain subregions in two dimensional conformal field theories (CFT) \cite{Calabrese:2005in, Calabrese:2006rx}. A related question is: to what extent is the dynamics of entanglement universal in non-relativistic scale-invariant theories in two (and higher) dimensions? The symmetry groups admitting anisotropic scale-invariance are not powerful enough to fix the dynamics. Despite this fact, we show that it is possible to learn important lessons about entanglement dynamics in (anisotropic) scale-invariant integrable theories, utilizing the celebrated quasi-particle (QP) picture \cite{Calabrese:2005in, Alba:2016}.

\begin{figure}[h!]
\begin{center}
\includegraphics{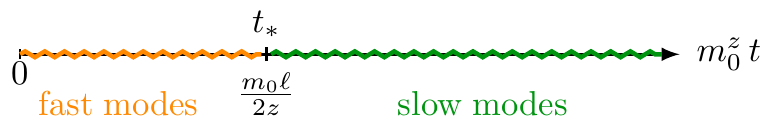}
\end{center}
\caption{After the quantum quench at $t=0$, before and after $t_*\equiv m_0^{1-z}\ell/2z$, the fast and the slow modes carry the entanglement respectively. For $z>1$ the role of the slow modes are dominant and vice versa for $z<1$.}
\label{fig:fastslow}
\end{figure}

Our main focus is on two dimensional theories with dispersion relation
\be\label{eq:dis}
\omega=k^z,
\ee
where $z$ is positive and $z\neq 1$ \cite{f0}. These theories are invariant under Lifshitz scaling, which are interesting partially due to the symmetry structure of quantum critical points \cite{Sachdev}. We study the dynamics of entanglement entropy (EE) and mutual information (MI) followed by a quantum quench. The only relevant scale in this problem is the one in the pre-quench state, which we denote it by $m_0$. This scale is basically identified with the parameter that we take it to vanishes after the quench. We consider $m_0$ to be finite in our analysis and all physical quantities are compared with this scale. In our analysis we denote the $k< m_0$ modes by \textit{slow modes} and the rest $k> m_0$ by \textit{fast modes}. The scope of these modes after the quantum quench are illustrated in FIG.\ref{fig:fastslow}.

In the following we mainly focus on the $z>1$ case in detail and briefly explain the corresponding differences with the $z<1$ case at the end.
\section{Dominance of the slow modes}
We consider integrable models with the dispersion relation \eqref{eq:dis} and use the QP picture, uplifted with the integrability knowledge of the final steady state \cite{Calabrese:2005in, Alba:2016} to understand the dynamics of entanglement. The EE of a connected interval of length $\ell$ is given by 
\begin{align}\label{eq:AC1}
S(t)
=2t\int\displaylimits_{2|v(k)|t<\ell}\, dk \,s(k)v(k)+\ell \int\displaylimits_{2|v(k)|t>\ell}\, dk \,s(k),
\end{align}
where $v(k)$ is the group velocity of the QPs given by $v(k)=\partial\omega/\partial k=z\,k^{z-1}$ and $s(k)$ denotes the individual contribution of modes with momentum $k$ to the entropy. We consider a more explicit form of \eqref{eq:AC1} as
\begin{align}\label{eq:AC2}
S(t)=2t\int_{0}^{k^*_\ell}\, dk \,s(k)v(k)+\ell \int_{k^*_\ell}^{\infty}\, dk \,s(k),
\end{align}
where
\be
k^*_\ell=\left(\frac{\ell}{2zt}\right)^\frac{1}{z-1}
\ee
is a characteristic momentum corresponding to time $t$, in which $k>k^*_\ell$ has been saturated before $t$ and $k<k^*_\ell$ are still contributing to the time evolution of EE. This $k^*_\ell$ is a decreasing function of time and $k^*_\ell(t_*)=m_0$. Thus using the terminology introduced in FIG.\ref{fig:fastslow}, for $t<t_*$ the fast modes only contribute to the dynamics and afterwards the slow modes take this role. The role of the slow modes stands until infinite time, though there is no (finite) saturation time in these theories as apposed to relativistic cases. 

Alba and Calabrese used the fact that in integrable theories, the state of the system will finally relax to a generalized Gibbs ensemble to fix the $s(k)$ in terms of the expectation value of the number operator in the pre-quench state as \cite{Alba:2017lvc}
\be
2\pi\,s(k)=-n_k\ln n_k\pm(1\pm n_k)\ln(1\pm n_k)\;,
\ee
where the upper and the lower signs correspond to bosonic and fermionic theories.

The analysis in this paper is quite general for theories with \eqref{eq:dis}.
In order to perform explicit calculations, we consider two family of bosonic and fermionic theories as the prototypes to study anisotropic scale-invariant fixed points. These theories are generalizations of Klein-Gordon and Dirac fermion theories defined as \cite{Alexandre:2011kr}
\begin{align}
\mathcal{S}_b&=\frac{1}{2}\int dt d\vec{x} \left[\dot{\phi}^2-\phi\left(\left(-\Delta\right)^z -m^{2z}\right)\phi\right]
\label{eq:ActionScalar},
\\
\mathcal{S}_f&=\frac{1}{2}\int dt d\vec{x}\,\bar{\Psi}\left(i\gamma^0\partial_t+i\Delta^{\frac{z-1}{2}}\gamma\cdot\partial-m^z\right)\Psi,
\label{eq:ActionFermion}
\end{align}
where the bosonic theory is defined for integer values of $z$ and the fermionic theory is defined for odd values of $z$ and both theories are invariant under Lifshitz scaling $(t,x)\to(\lambda^zt,\lambda x)$ when $m\to0$. We explicitly study these theories in two dimensional spacetime, though the analysis is generalizable to higher dimensions for spherically symmetric entangling regions \cite{longversion}.
\begin{figure*}[t!]
\begin{center}
\includegraphics[scale=0.31]{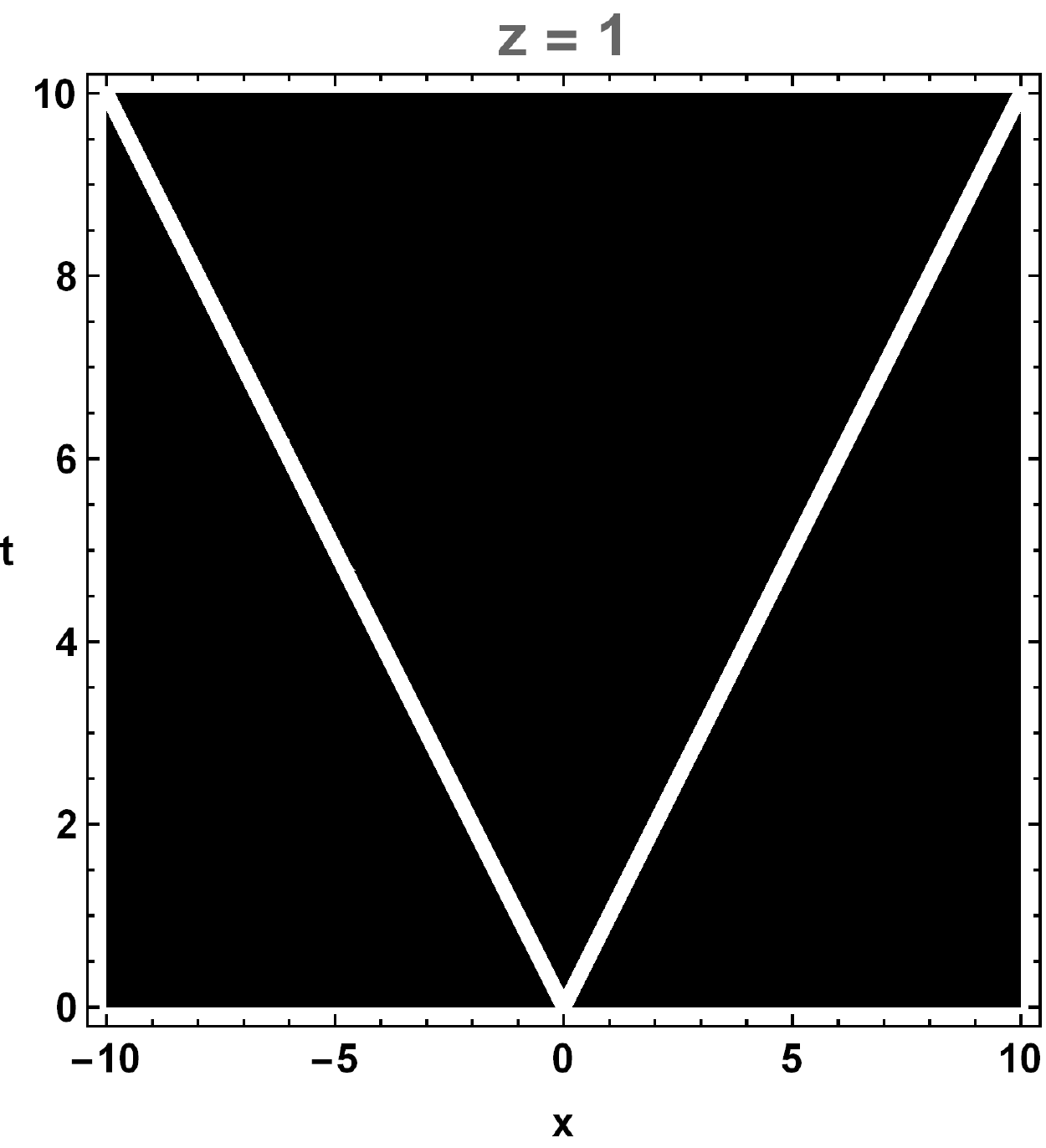}
\includegraphics[scale=0.3]{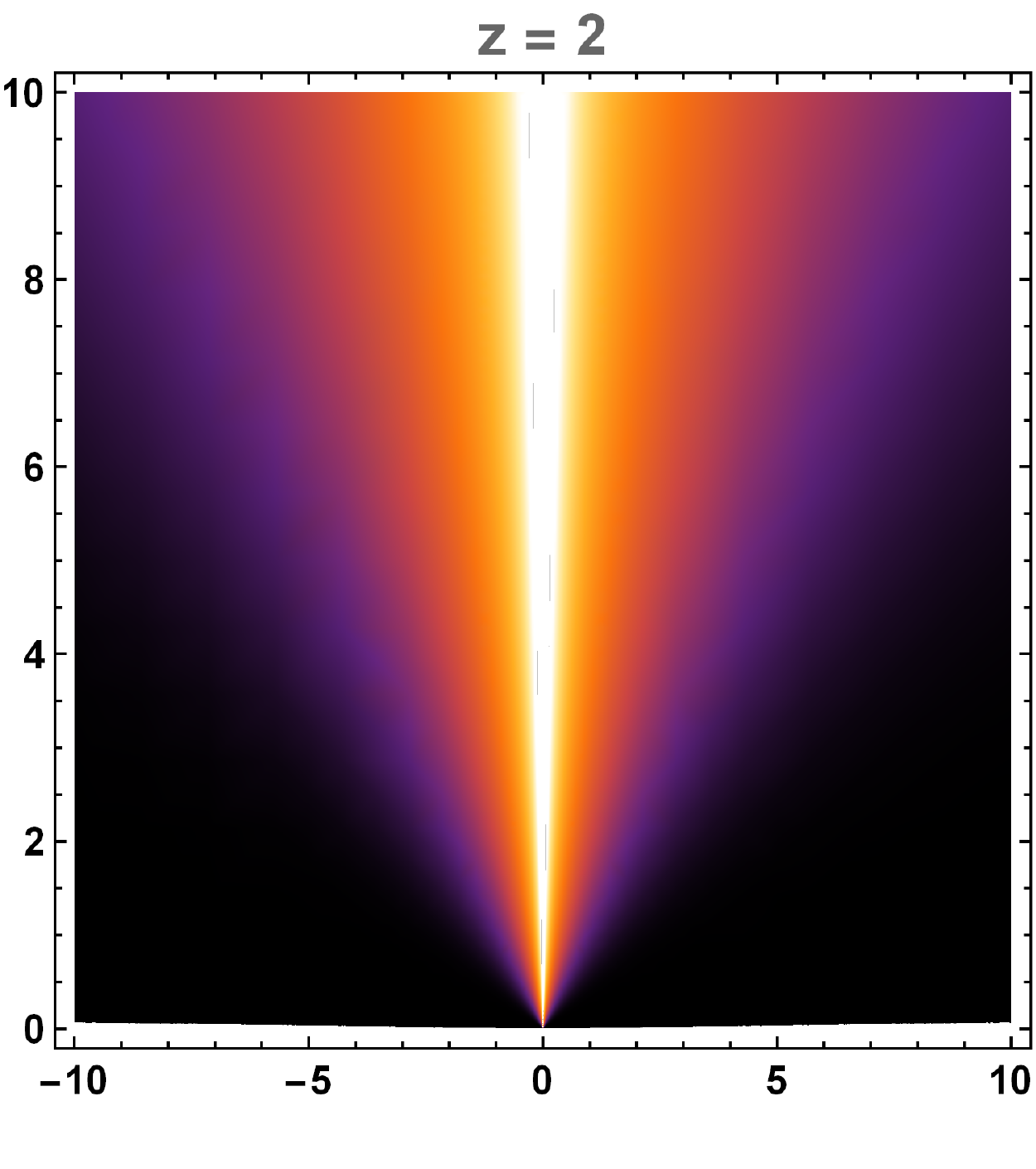}
\includegraphics[scale=0.3]{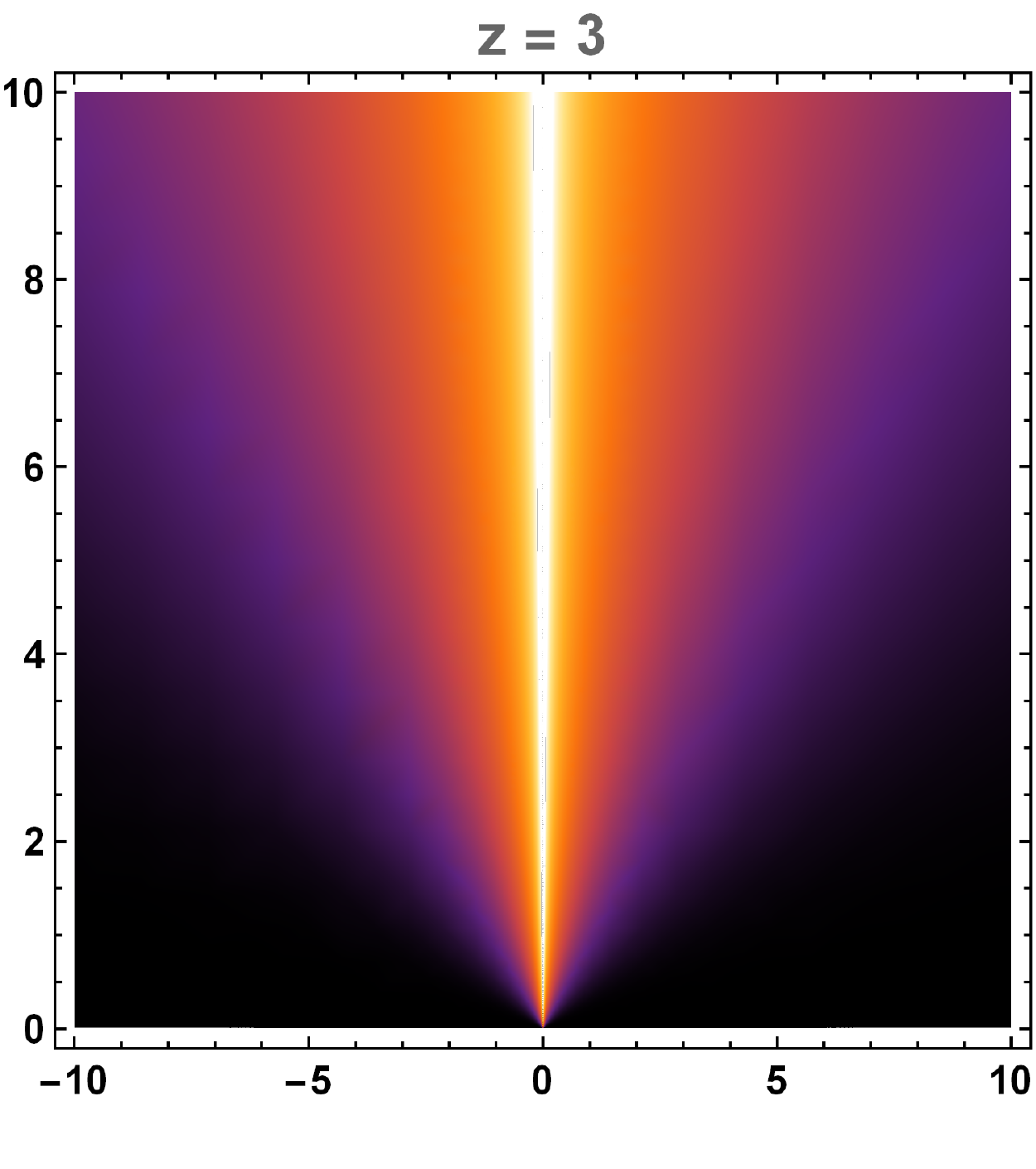}
\includegraphics[scale=0.3]{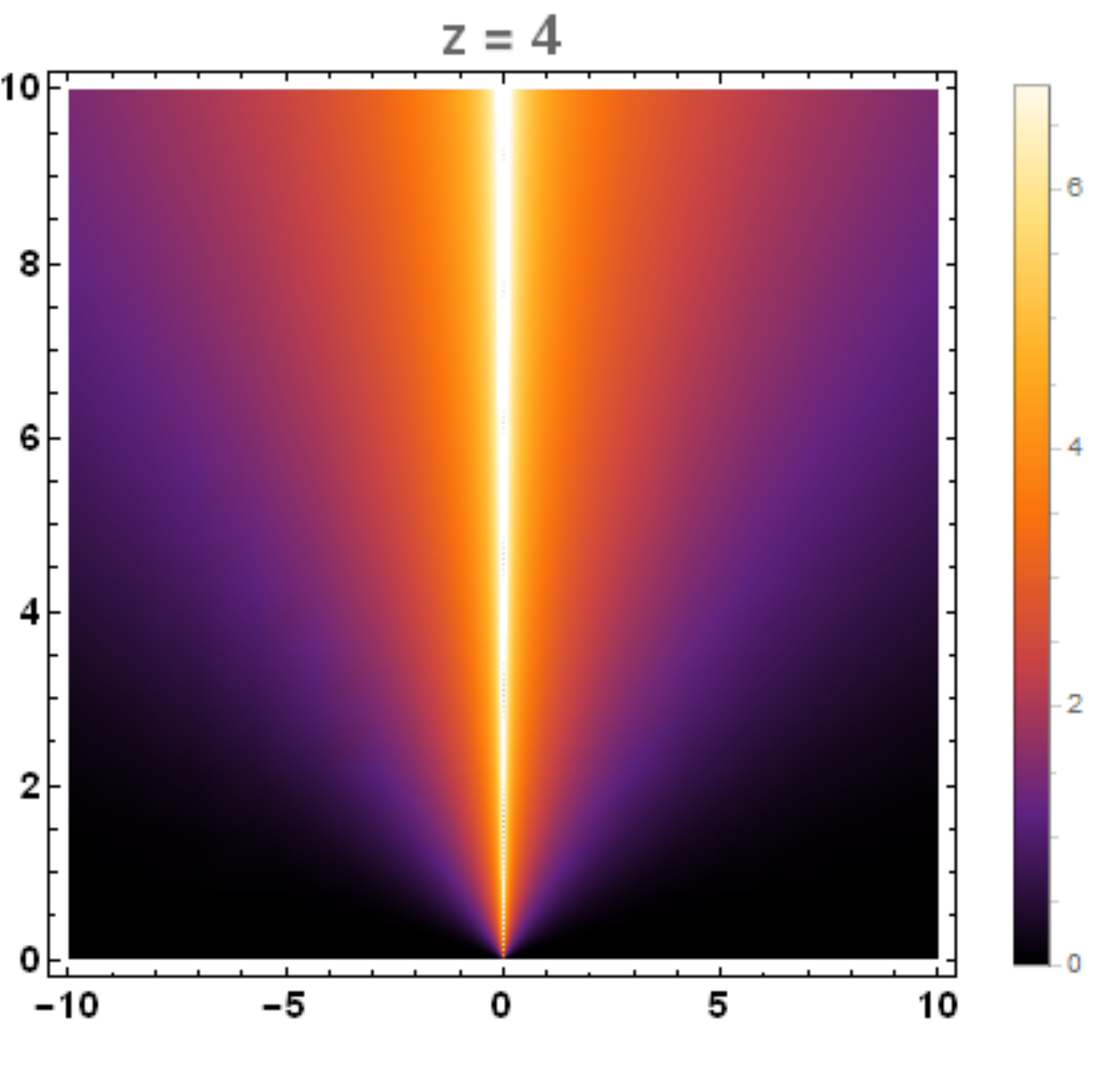}
\end{center}
\caption{Density plot of $s(k)$ for bosonic theory on the $t-x$ plane for $z=1,2,3,4$ from left to right. The $z=1$ case is obviously different from others and there exists a single mode with a constant (unit) group velocity. For $z>1$, $s(k)$ of the slow modes is large while it almost vanishes for the fast modes. The pre-quench dispersion relation is $\omega_0=\sqrt{m_0^{2z}+k^{2z}}$ and the post-quench dispersion is $\omega=k^z$ and in all plots we have set $m_0=1$. Fermionic theories have the same behavior in a squeezed scale.}
\label{fig:sk}
\end{figure*}

\section{Dynamics of Entanglement Entropy}
As a warm up, $s(k)$ is depicted in these theories in FIG.\ref{fig:sk}.
Although these non-relativistic theories are conceptually different from their relativistic counterparts, in which an upper bound exists on the group velocity of the propagating modes \cite{Calabrese:2005in, Casini:2015zua},
these density plots show that $s(k)$ enjoys an effective light-cone structure. The entropy is expected to be dominantly affected by the slow modes, populated close to the time axis \cite{fLR}. In the following we explicitly analyse the role of fast and slow modes separately and confirm our analysis with numerical checks.
\subsection{Fast modes}
The fast modes contribute to the entropy in the early steps of the evolution, namely for $t < t_*$. The larger is the value of $z$, the velocity of all physical modes is increased, thus they saturate earlier and this regime becomes shortened. In other words $t_*$ decreases for larger values of $z$. In this regime, $s(k)$ is given by
\begin{align}\label{eq:SkSeriesFast}
\begin{split}
2\pi\, s_b^\mathrm{fast}(k)&=\frac{m^{4z}_0}{16k^{4z}}\left(1-\ln \frac{m^{4z}_0}{16k^{4z}}\right)
+\mathcal{O}\left(\frac{m_0^{6z}}{k^{6z}}\right),
\\
2\pi\, s_f^\mathrm{fast}(k)&=\frac{4m^{2z}_0}{k^{2z}}\left(1-\ln \frac{m^{2z}_0}{4k^{2z}}\right)
+\mathcal{O}\left(\frac{m_0^{4z}}{k^{4z}}\right),
\end{split}
\end{align}
which approximates $s(k)$ for $k>M_\mathrm{f}\gtrsim m_0$. As we take into account higher orders of \eqref{eq:SkSeriesFast}, $M_\mathrm{f}\to m_0$. EE in this regime can be found by plugging \eqref{eq:SkSeriesFast} into \eqref{eq:AC2}, where the lower bound of the first integral is replaced by $M_\mathrm{f}$. The crucial point is that for $t\ll t_*$, we have $k_*\gg m_0$, and \textit{all} of the fast modes contribute to the linear growth of the EE through the lower bound of the first integral. Based on this considering infinite order of \eqref{eq:SkSeriesFast} leads to
\begin{align}\label{eq:slin}
\begin{split}
S^{\mathrm{fast}}_{b/f}(t)&=
\mathfrak{c}_{b/f} \,m_0^z\,t+\cdots,
\end{split}
\end{align}
where $\mathfrak{c}_b=(\pi-2)/4$ and $\mathfrak{c}_f=1/2$ for bosonic and fermionic theories. Note that these coefficients are \textit{universal} in the sense that they are \textit{independent} of $z$ for $z>1$ and their value is equal to the corresponding relativistic counterparts (see for instance \cite{Cotler:2016acd}). This universality was expected since the structure of the fast modes in $s(k)$ does not very much depend on $z$, though one should be careful that the regime of validity of this linear scaling is \textit{shortened} by a factor of $1/z$.
\begin{figure*}[t!]
\begin{center}
\includegraphics[scale=0.30]{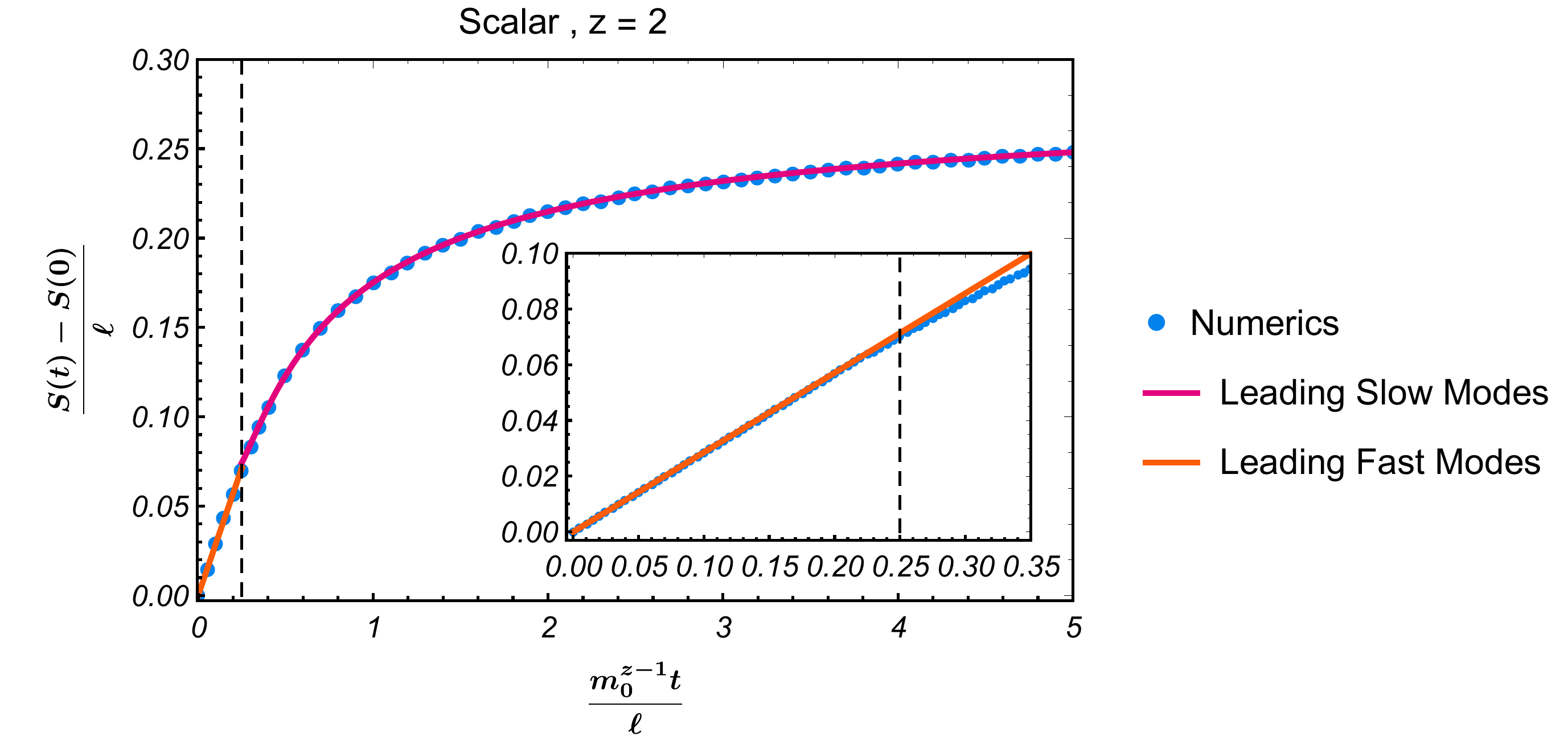}
\hspace{5mm}
\includegraphics[scale=0.30]{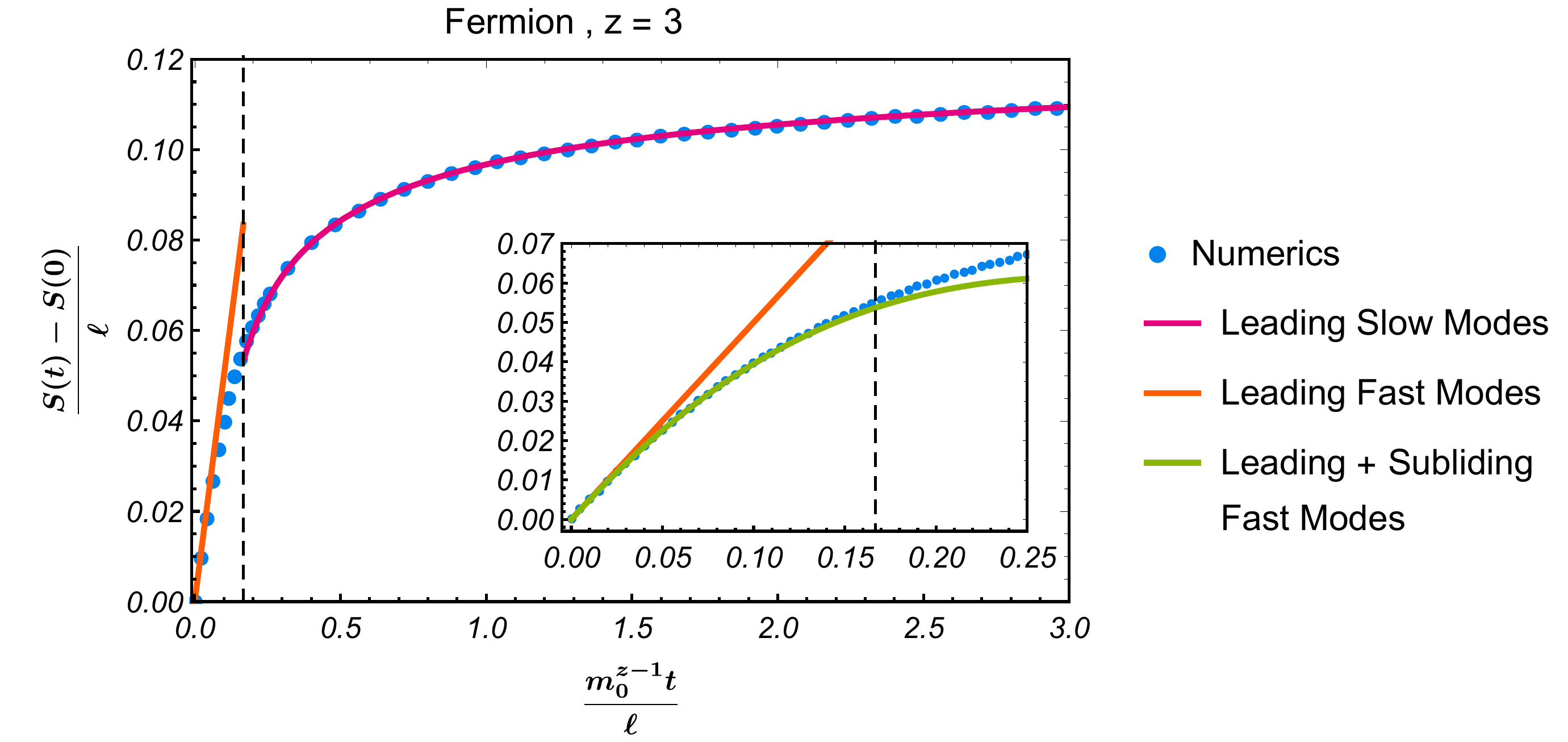}
\end{center}
\caption{QP prediction for the evolution of EE in scalar and fermionic theories. The dashed vertical line corresponds to $t=t_*$. The inner panels show the same graph with focus on $t<t_*$. We have set $m_0=1$ in both plots and consider $-\infty<k<+\infty$. We set $M_b=0.7$ and $M_f=0.95$. In the fast mode regime of the scalar theory, the linear term well approximates almost all the $t<t_*$ region though in the fermionic theory we need to add at least one higher order correction. The structure of these plots is the same for any higher values of $z$.
}
\label{fig:QPapprox}
\end{figure*}

The structure of the ellipsis in \eqref{eq:slin} is given by $\mathfrak{m}_0^{2nz}\,\mathfrak{t}^\frac{1-2nz}{1-z}$ where $\mathfrak{t}\equiv t/\ell^z$,  $\mathfrak{m}_0\equiv m_0\ell$ and $n=1,2,\cdots$ stands for the expansion order. The scaling of these terms are universal but such terms are suppressed with the factor of $\ell^{\frac{z-2 n z}{z-1}}$. The first order correction in fermionic theories is given by
\begin{align}
\begin{split}
S_{f,1}^{\mathrm{fast}}(t)&=
\frac{2z^2 \alpha_z \mathfrak{m}_0^{2 z}}{(1-2 z)}  \mathfrak{t}^{\frac{1-2 z}{1-z}} \left[\mathfrak{f}_1
-z \ln \left(\frac{ 2^{\frac{1}{z}}z\,\mathfrak{t}}{\mathfrak{m}_0^{1-z}}
\right)
\right]\;,
\end{split}    
\end{align}
where
$\mathfrak{f}_1=(1-z) (8 z-3)/2(1-2 z)$ and $\alpha_z=(2 z)^{\frac{1}{1-z}}/(2\pi\,z)$. The same order of correction vanishes in bosonic theories due to the structure of $s(k)$.
These subleading terms (mixed up with their counterparts among the slow modes) provide a smooth transition between these two regimes around $t \sim t_*$.
We have shown the validity of this approximation and the transition in FIG.\ref{fig:QPapprox}.

\subsection{Slow Modes}
The effect of the slow modes \textit{drastically} changes the story of entanglement propagation in theories with $z>1$ compared to relativistic scale-invariant theories, $z=1$. Due to the comparatively large $s(k)$ of the slow modes, they carry most of the entanglement in these theories, starting to contribute from $t =t_*$ and standing until infinite time. To analyse the contribution of the slow modes, we consider $s(k)$ in this regime which is given by
\begin{align}\label{eq:SkSeriesSlow}
\begin{split}
2\pi\,s_b^\mathrm{slow}(k)&=\ln \frac{m^{z}_0}{k^{z}}+\left(1-\ln 4\right)
+\mathcal{O}\left(\frac{k^{2z}}{m_0^{2z}}\right)\;,
\\
2\pi\,s_f^\mathrm{slow}(k)&=\ln 2-\frac{k^{2z}}{2m_0^{2z}}+\mathcal{O}\left(\frac{k^{4z}}{m_0^{4z}}\right)\;,
\end{split}
\end{align}
where these expansions well-approximate $s(k)$ for $k<M_{\mathrm{b}}\lesssim m_0$. This well approximation is due to the structure of $s(k)$ in these theories, namely the high population of very slow modes. By plugging these expressions into \eqref{eq:AC2}, where the upper bound of the second integral has been replaced by $M_{\mathrm{b}}$, the EE read
\begin{align}\label{eq:EEslow}
\begin{split}
S^{\mathrm{slow}}_b(t)&=
\mathfrak{b}_1\,\ell \,M_{\mathrm{b}}
+\alpha_z \,\mathfrak{t}^{\frac{1}{1-z}} \left[\mathfrak{b}_2-z \ln \left(\frac{2z\, \mathfrak{t}}{ \mathfrak{m}_0^{1-z}}\right)\right]+\cdots,
\\
S^{\mathrm{slow}}_f(t)&=
\frac{\ln 2}{2\pi}\, \ell \,M_{\mathrm{b}} +\ln 2\,\alpha_z(1-z)\,\mathfrak{t}^{\frac{1}{1-z}}+\cdots,
\end{split}
\end{align}
where $2\pi\,\mathfrak{b}_1=\ln(m_0^z e^{z+1}/4M_{\mathrm{b}}^z)$ and $\mathfrak{b}_2=1-z^2+(z-1)(\ln 4-1)$.
This $\mathfrak{t}^{\frac{1}{1-z}}$ function presents the scaling of EE during most of the evolution, namely $t_*<t<\infty$. Note that the logarithmic term in the bosonic theories is originated from the logarithmic divergence of $s(k)$ of the very slow modes in these theories \cite{fNL}.

\begin{figure*}[t!]
\begin{center}
\includegraphics[scale=0.30]{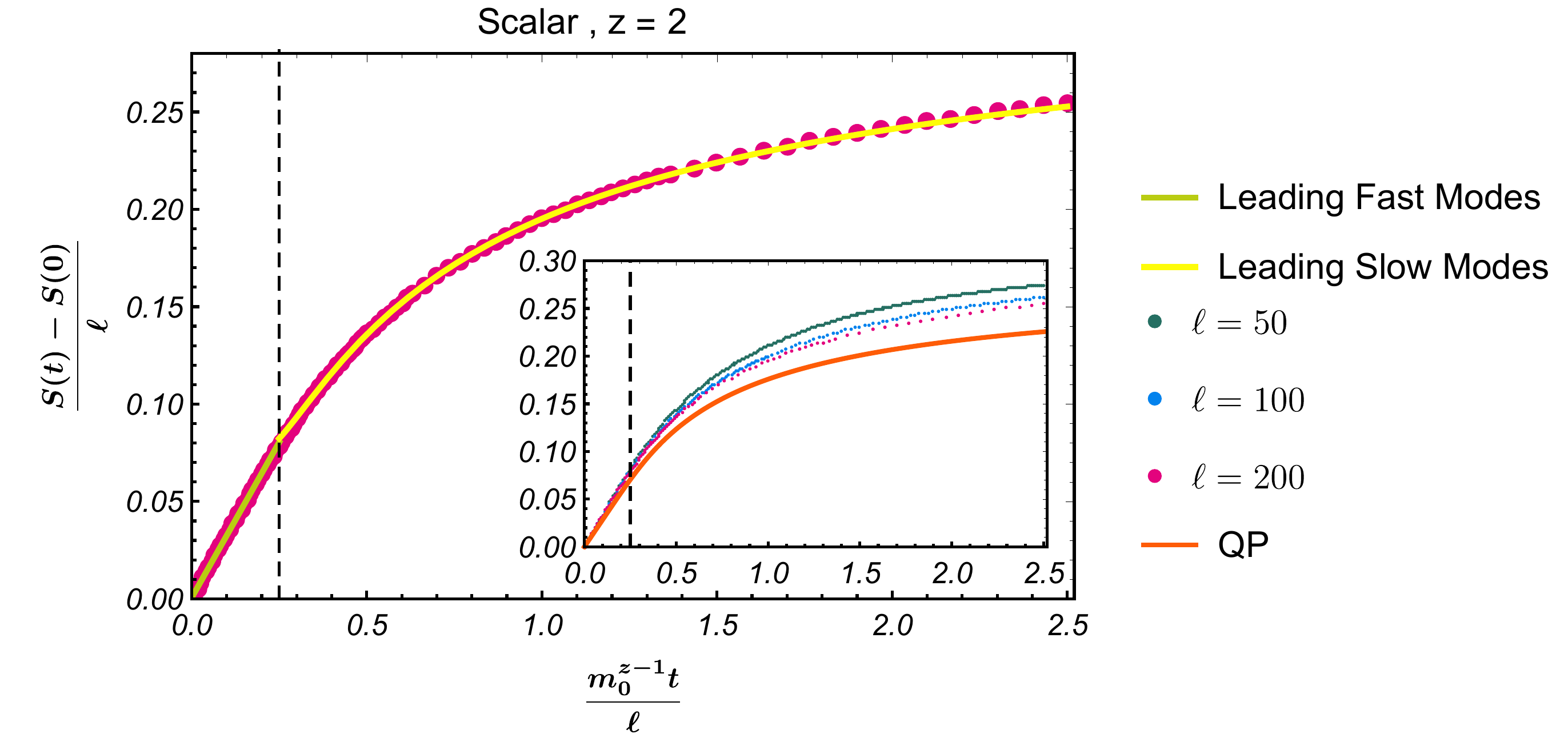}
\hspace{5mm}
\includegraphics[scale=0.30]{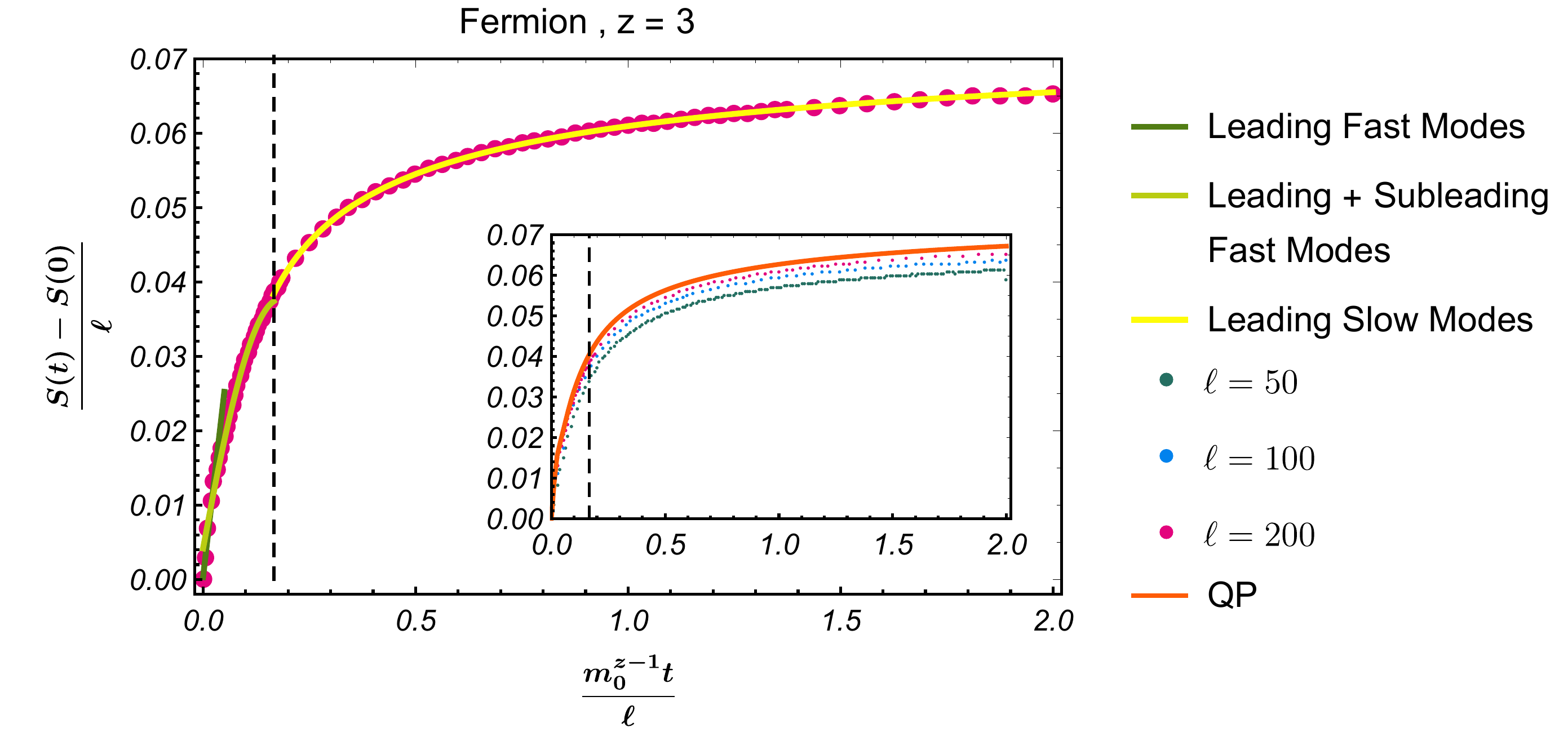}
\end{center}
\caption{Numerical results for the evolution of EE in scalar and fermionic theories. The inner panels show how numerics approach QP prediction as the system size is increased. We have set $m_0=1/2$ in the right panel and $m_0=1/4$ in the left panel. We have used an IR cutoff $m=10^{-5}$ for the scalar theory. These numerics are found on a lattice where $-1<k<+1$. The fitted functions for the scalar case are $0.32\,t$ (green) and $0.32-t^{-1}(0.125+0.047\ln t)$ (yellow). The fitted functions for the fermionic case are $0.31\,t-1.59\,t^{5/2}$ (green) and $0.076-0.016\,t^{-1/2}$ (yellow).}
\label{fig:LatticeNumrics}
\end{figure*}
In FIG.\ref{fig:QPapprox} we show how these first order approximations \eqref{eq:EEslow} lie on the top of the exact (numerically found) values of \eqref{eq:AC2}. This well approximation in bosonic models is due to the blow up of $s(k)$ for the very slow modes, and in the fermionic models is due to the slowly varying nature of $s(k)$ in the slow mode regime. Though the first order approximation works quit well, it is straightforward to work out the higher orders which their structure is as $\mathfrak{m}_0^{2nz}\,\mathfrak{t}^\frac{1+2nz}{1-z}$ with $n=1,2,\cdots$. The exact structure of the entropy around $t\approx t_*$ is complicated due to the mixture of the subleading effects of both regimes. 

Before getting into the numerical checks, we would like to interpret our expressions for the case of $z<1$. In this case as $z$ is decreased starting from $z=1$: a) $t_*$ increases and the fast modes become more involved in the dynamics, b) the density plots in FIG.\ref{fig:sk} become more uniformized. This weaker fall-off of $s(k)$ for large momenta, accompanied by a weaker blowup for small momenta in bosonic case, ends up with an increase in the contribution of $\mathfrak{m}_0^{2nz}\,\mathfrak{t}^\frac{1-2nz}{1-z}$ terms (in the fast mode regime) and reaching a sharp saturation at $t_*$.

\subsection{Numerical Results}
In this part we report how numerical results for the EE in the vacuum state of the bosonic and fermionic theories match with the aforementioned analytic predictions. To find these numerical results we use the correlation matrix method for Gaussian states \cite{Cor} to study quantum quenches from a massive theory to a scale-invariant theory. We consider regularized versions of \eqref{eq:ActionScalar} and \eqref{eq:ActionFermion} on an infinite lattice given by (see \cite{LHLM} and the appendix of this paper)
\begin{align}
H_b&=\frac{1}{2}\sum_n\left[\pi_n^2+\left(\partial^z\phi_{n}\right)^2+m^{2z}\phi_n^2\right],
\\
H_f&=\sum_n\left[-\frac{i}{2}\left(\Psi^\dagger_n\gamma^0\gamma^1\partial^z\Psi_{n}-\mathrm{h.c.}\right)+m^z\Psi^\dagger_n\gamma^0\Psi_{n}\right],
\end{align}
where $\partial^z\,f_{n}=\sum_{k=0}^z(-1)^{z-k}{}_zC_k\,f_{n+k}$ and ${}_zC_k$ is the binomial coefficient \cite{f2}. Numerical results corresponding to a single interval are presented in FIG.\ref{fig:LatticeNumrics}. We find a very good agreement between numerical results and the QP predictions for a much wider family of parameters which we have not presented here. The time scaling of EE after boundary state quench as well as entropy production in bosonic and fermionic Gaussian thermofield double states have been also found to be in a very good agreement with the QP predictions \cite{longversion}.

Aside from the numerical checks, we can justify the scaling behaviour corresponding to the slow modes with a direct calculation of the spectrum of $\rho_A$ in a semi-analytic way. For instance in bosonic theories we numerically find that, \textit{at most}, the first $z$ largest eigenvalues scale with time as a power law, while the rest of them scale logarithmically. Using the correlator method for very small subregions, we can analytically find the exponents of these dominant eigenvalues. Assuming that the coefficient of the leading power law is a slow varying function among these dominant eigenvalues, integrating over these eigenvalues in the continuum recovers the aforementioned $\mathfrak{t}^{\frac{1}{1-z}}$ scaling of the EE \cite{longversion}.    

\section{An Implication on Scrambling of Local Quantum Information}
MI of separated regions is an important correlation measure which its dynamics quantifies how local quantum information scrambles (spreads) over larger subregions \cite{Alba:2019ybw, f1}. In CFTs, although MI generally depends on the full spectrum of the theory, the pattern of its time evolution is well-known. Putting aside the cases of very large central charges where the QP picture fails \cite{chaoticCFTs}, in these theories (even more generally in any integrable theory that most of the entanglement is carried by the fastest QP),  MI exhibits a peak at some finite time. More precisely, when there is an upper bound on the QP velocities, MI starts to raise after a certain time and peaks at $t_p=(d+\ell)/(2v_m)$, where $v_m$ is the velocity of the fastest mode, $\ell$ denote the subregions size and $d$ is the separation between them. Moreover, MI starts to decay to zero after this peak.

It may seem that for $z>1$, since there is no upper bound on the velocity of the propagating modes, MI should instantly peak and then start to decay. As we have shown in the previous section, the slow modes carry most of the entanglement, so the story is different with mostly known cases including CFTs where $z=1$. In theories obeying \eqref{eq:dis}, the MI starts to grow very slowly right after the quench (due to the very fast modes which carry a tiny amount of entanglement) and smoothly starts to raise significantly after the slower modes start to contribute. There is a peak due to the effect of the slow modes, and afterwards the peak decays \textit{slowly}.

\begin{figure}
\begin{center}
\includegraphics[scale=0.31]{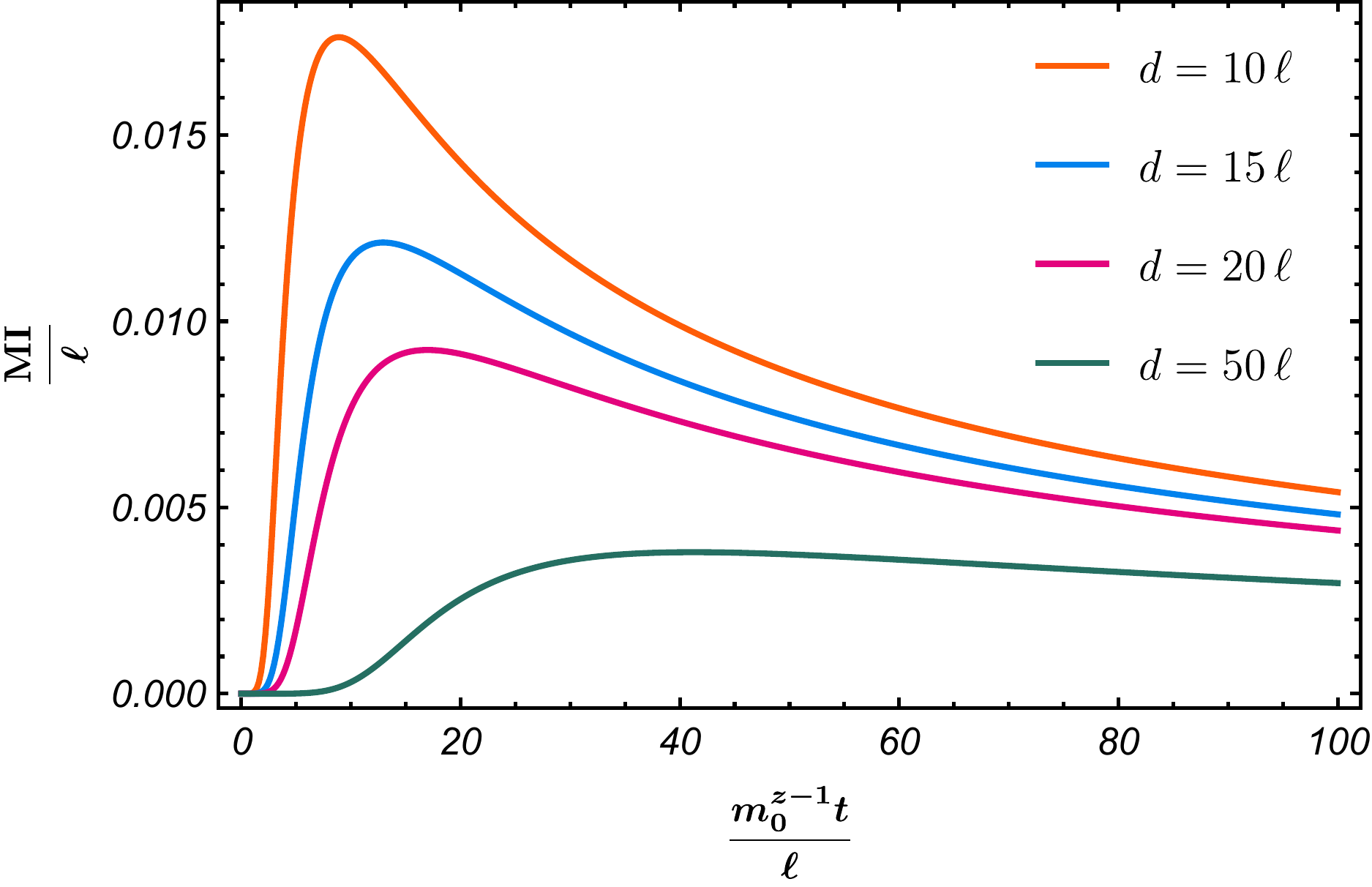}
\includegraphics[scale=0.3]{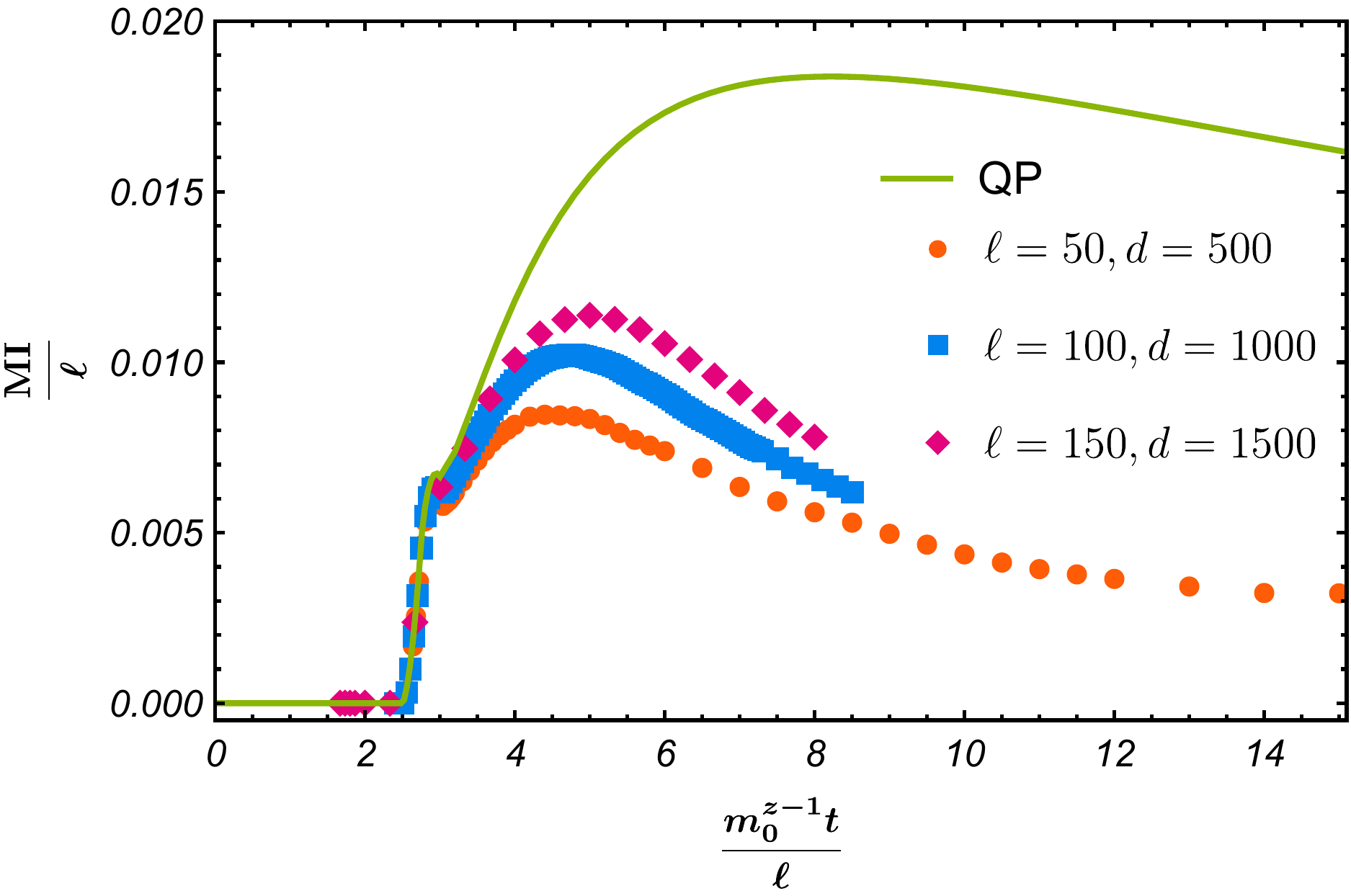}
\includegraphics[scale=0.32]{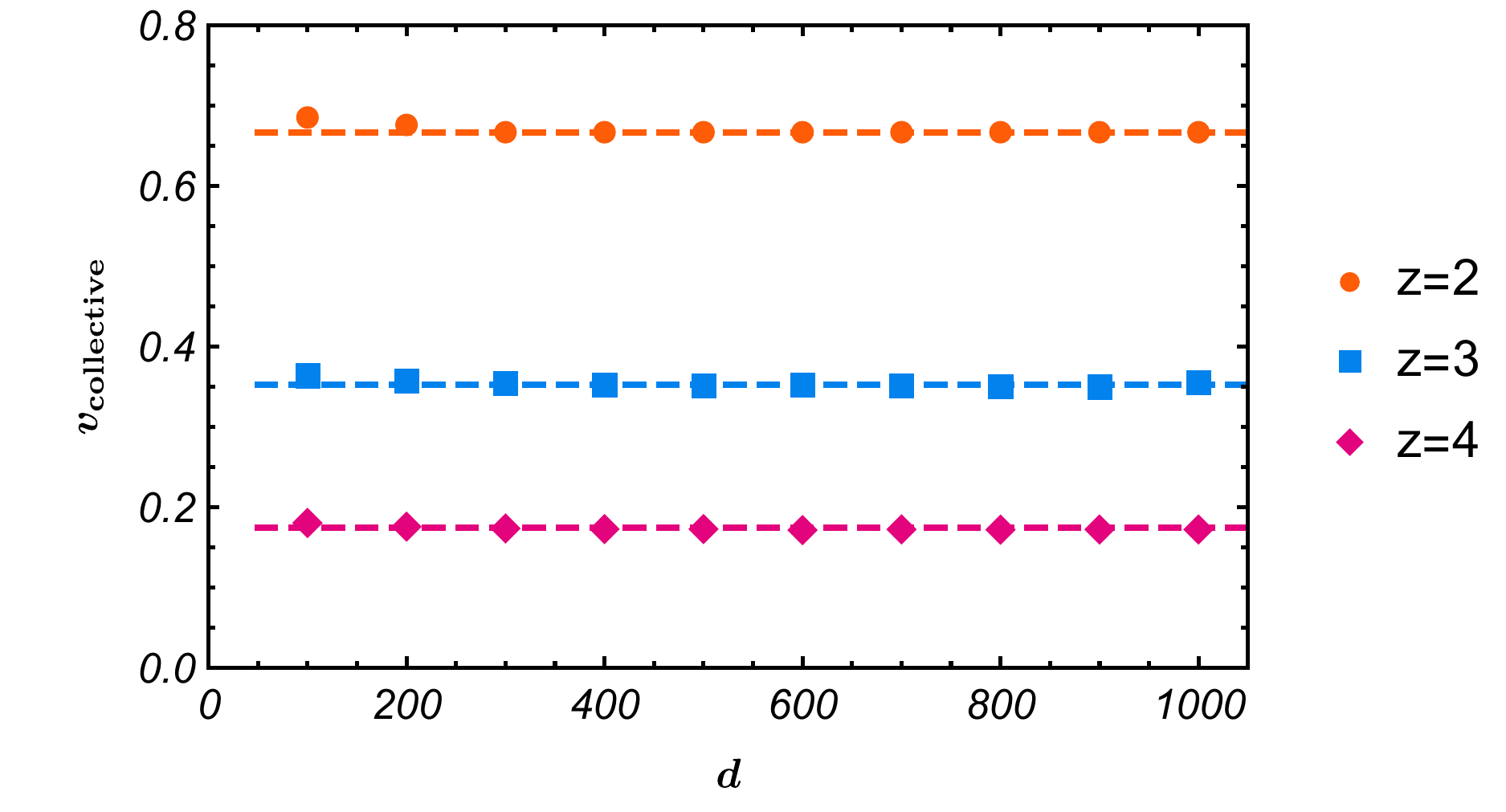}
\end{center}
\caption{MI for scalar theory (the structure is very similar for fermionic theories). The left and the middle panels correspond to $z=2$. The left panel shows QP prediction for the existence of the peak and how it decays with $d$. In the middle we show the agreement between numerical calculations and QP analysis. We show how the numerics approach to the QP prediction in $\ell\to\infty$ for $d/\ell=10$. The tiny peak slightly before the horizontal axis takes $\sim$ 3 is a lattice effect which will be suppressed for larger separations \cite{MohammadiMozaffar:2018vmk}. In the right panel we show how the peak corresponding to the slow modes can be attributed to a collective mode which its velocity is defined as $v_\mathrm{collective}=(d+\ell)/(2t_p)$. All these $z>1$ collective velocities are $<1$ (corresponding to $z=1$).}
\label{fig:MI}
\end{figure}

We analyse the dynamics of MI similar to the previous section. The expression counterpart to \eqref{eq:AC1} for MI was introduced in \cite{Alba:2016}, and in the same manner that we wrote \eqref{eq:AC2}, we can rewrite MI as, 
\begin{align}
\begin{split}
I(t)
&=
\int_{k^*_d}^{k^*_{d+\ell}}\, dk\, s(k)\left(vt-\frac{d}{2}\right)
+\int_{k^*_{d+\ell}}^{k^*_{d+2\ell}}\, dk\, s(k)\left(\frac{d}{2}+\ell-vt\right).
\end{split}
\end{align}
Since the whole resulting expression may not be informative, we only present it for far apart subregions, namely $d\gg\ell$, where the peak of MI occurs at
$m_0^z\,t_p\approx\mathfrak{r}^{z-1}_{b/f}\frac{m_0\,d}{2z}$,
where $\mathfrak{r}_{b}=\left(4 e^{z-1}\right)^{\frac{1}{z}}$ and $\mathfrak{r}_{f}=\left(\frac{2 z+1}{\ln 4}\right)^{\frac{1}{2 z}}$. MI at the peak is given by 
\be
I(t_p)\approx\mathfrak{g}_{b/f}\frac{z}{(z-1)}\frac{\ell^2\,m_0}{d},
\ee
where $\mathfrak{g}_{b}=1/\mathfrak{r}_{b}$ and $\mathfrak{g}_{f}=1/\mathfrak{r}_{f}^{2z+1}$. The MI peak decays with $d^{-1}$, which is stronger that the relativistic case decaying with $d^{-\frac{1}{2}}$, due to the dominance of the fast modes \cite{Alba:2019ybw}. Interestingly this scaling is \textit{independent} of the dynamical exponent, \textit{i.e.},  universal over all $z>1$ theories. In FIG.\ref{fig:MI} we have shown the evolution of MI predicted by the QP picture and confirmed numerically. 
The decay of the MI itself is also universal in these theories given by
\begin{align}
I_f(t>t_p)&\approx
\mathfrak{i}_{b/f}\frac{z\,\alpha_z}{(z-1)}
\frac{\ell ^2}{d^2}
\left(\frac{t}{d^z}\right)^{\frac{1}{1-z}},
\end{align}
where $\mathfrak{i}_{b}=1$ and $\mathfrak{i}_{f}=\ln 2$. The bosonic case is again enhanced with a logarithmic correction as well.

We would like to also comment that entanglement revivals are also able to capture scrambling of local quantum information into global degrees of freedom \cite{Modak:2020faf}. By putting these theories on compact spatial dimension, we find very similar result to MI for the shortening of the deep of entanglement revival of a connected interval which will reported in future work \cite{longversion}.     

\section{Conclusions and discussions}
We analysed the propagation of entanglement in integrable scale-invariant theories. We showed that the scale of the pre-quenched Hamiltonian divides the dynamics into distinct regimes. Most of the dynamics of entanglement is understood in terms of the slow modes with a certain time scaling that stands until infinite time. We showed that this feature causes a universal scrambling of local quantum information in these theories stronger than relativistic theories.

It is worth to note that there is a detailed literature behind correlation and entanglement dynamics in long-range interacting models (see e.g. \cite{Tagliacozzo:2013}). These models admit strict Lieb-Robinson bounds \cite{LLC}, though certain experiments verify the existence of propagating modes which violate Lieb-Robinson bounds \cite{nature12}. The scale-invariant theories studied here are similar to this family of models and our theoretical explanation is in complete agreement with the aforementioned experiments.

We would like to also note that holographic studies in Lifshitz $z>1$ theories results in linear growth and sharp saturation (after a finite time) of EE \cite{holographicResults}. Assuming the universality of the results of this paper among anisotropic scale-invariant theories, the holographic results seem to be seriously questionable. Considering this problem in a more general sense, including the entanglement structure in static states, there also exists other important observations from direct field theory calculations of EE at Lifshitz fixed-points which does not agree with the corresponding holographic results. Most importantly, the field theory results are clearly $z$-dependent while the holographic results are $z$-independent. We believe that this disagreement originates in the relativistic nature of the underlying theory in the so far studied holographic models, which results in conceptual differences, for instance in the causal structure of the holographic dual theory, in comparison with field theory constructions of Lifshitz theories.

As already mentioned, the strong role of the slow modes in these theories is tempting the existence of Lieb-Robinson bounds. Proving such a bound would be a very interesting future direction. Another interesting direction to explore is to analyze the dynamics of EE near anisotropic quantum critical points (see \cite{Castro-Alvaredo:2020mzq} for a recent related study).

\subsection*{Acknowledgements}
We thank Masahiro Nozaki and Tadashi Takayanagi for useful discussions. We specially thank Vincenzo Alba, Pasquale Calabrese and Erik Tonni for carefully reading this manuscript and their fruitful comments. AM is supported by JSPS Grant-in-Aid for Challenging Research (Exploratory) 18K18766. AM was supported by Alexander von Humboldt foundation during the early stages of this project.

\pagebreak
\widetext
\begin{center}
\textbf{\large Appendix: Details of Numerical Models}
\end{center}
\setcounter{equation}{0}
\setcounter{figure}{0}
\setcounter{table}{0}
\makeatletter

\section{Fermionic Model}
In this appendix we first review how the relativistic Dirac fermion theory can be considered on a lattice and introduce a family of models as non-relativistic generalizations of Dirac fermion with dynamical exponent $z$. 
\subsection{Relativistic Dirac Fermion}
We consider the Lagrangian density given by
\be
\mathcal{L}=\frac{1}{2}\bar{\Psi}(i\gamma^\mu\partial_\mu-m)\Psi\;.
\ee
The discrete Hamiltonian on an infinite lattice is given by
\be
H=\sum_n\left[-\frac{i}{2}\left(\Psi^\dagger_n\gamma^0\gamma^1\left(\Psi_{n+1}-\Psi_{n}\right)-\mathrm{h.c.}\right)+m\Psi^\dagger_n\gamma^0\Psi_{n}\right]\;,
\ee
where
$\{\Psi^{\dagger}_r,\Psi_s\}=\delta_{rs}$
and the $\gamma$ matricies satisfy Clifford algebra. Plugging the Fourier expansion as
\be
\Psi_n=\frac{1}{\sqrt{2\pi}}\int_{-\pi}^{\pi}dk\, \psi_k\, e^{-i k n}
\ee
into the Hamiltonian leads to
\be
H=\sum_n \psi^\dagger_k\left(-\gamma^0\gamma^1\sin k+m\gamma^0\right)\psi_{k}\;.
\ee
In order to diagonalize the model we treat the two components of the field separately as
\be
\psi_k=
\begin{pmatrix}
u_k\\ d_k
\end{pmatrix}
\ee
and use the basis $\gamma^0=\sigma_1$ and $\gamma^1=i\sigma_2$ which leads to
\be
H=\int_{-\pi}^{\pi}dk\, \left[\left(u^\dagger_k u_k-d^\dagger_k d_k\right)\sin k+m\left(u^\dagger_k d_k+d^\dagger_k u_k\right)\right]\;.
\ee
We consider the following Bogoliubov transformations 
\be\label{eq:Bog1}
u_k=\cos\frac{\theta_k}{2}\, b_k+i\sin\frac{\theta_k}{2}\, b^\dagger_{-k}
\;\;\;\;,\;\;\;\;
d_k=\sin\frac{\theta_k}{2}\, b_k-i\cos\frac{\theta_k}{2}\, b^\dagger_{-k},
\ee
where
\be
\tan\theta_k=\frac{m}{\sin k}
\;\;\;\;\;,\;\;\;\;\;
\cos\theta_k=\frac{\sin k}{\sqrt{m^2+\sin^2 k}}\;,
\ee
and $\{b^\dagger_k,b_{k'}\}=\delta_{kk'}$, to bring the Hamiltonian into its diagonal form given by
\be
H=\int_{-\pi}^{\pi}dk\, \sqrt{m^2+\sin^2 k}\left(b_{-k}^\dagger b_{-k}+b_k^\dagger b_k\right)\;.
\ee
\subsection{Non-relativistic Dirac Fermion}
To construct similar scale-invariant models we use the following expression for the higher order spatial derivatives given by
$$\partial^z\Psi_n=\sum_{i=0}^z (-1)^{z-i}\;{}_zC_i \;\Psi_{n+i}\;,$$
where ${}_zC_i=\frac{z!}{(z-i)!i!}$ and $z$ can take any odd positive integer value, together with the same Bogoliobuv transformations used for the relativistic case in \eqref{eq:Bog1} where
\be
\tan\theta_k=\frac{m^z}{f_z(k)}
\;\;\;\;\;,\;\;\;\;\;
\cos\theta_k=\frac{f_z(k)}{\Omega_k}\;.
\ee
So the diagonalized Hamiltonian is given by
\be
H=\int_{-\pi}^{\pi}dk\, \Omega_k\left(b_{-k}^\dagger b_{-k}+b_k^\dagger b_k\right),
\ee
where
\be
\Omega_k^2=m^{2z}+f_z(k)^2
\;\;\;\;\;\;,\;\;\;\;\;
f_z(k)^2=
\left(2\sin\frac{k}{2}\right)^{2z}\cos^2\frac{z k}{2}\;.
\ee
\subsection{Entanglement Entropy Following a Quantum Quench}
The vacuum correlator of the aforementioned model is given by
\be
\langle \Psi^\dagger_r\Psi_s\rangle=\frac{\delta_{rs}}{2}\mathbf{1}-\frac{1}{4\pi}\int_{-\pi}^{\pi}\frac{dk}{\Omega_k}
\begin{pmatrix}
f_z(k) & m^z\\ m^z & -f_z(k)
\end{pmatrix}
e^{i(r-s)k}.
\ee
In order to work out the entanglement entropy using the correlator method, the two point functions of the time evolved state is needed. We consider the initial state to be the vacuum state of the aforementioned Hamiltonian, which we denote its dispersion relation with $\Omega_{0,k}$. After the quench the state is evolved with $e^{-i\Omega_{k}t}$ 
that it is not hard to show that the components of the two point function $\langle \Psi^\dagger_r\Psi_s\rangle$ for the post-quench state are given by 
\begin{align}
\begin{split}
\langle u^\dagger_r u_s \rangle
&=\frac{\delta_{rs}}{2}-\frac{1}{4\pi}\int_{-\pi}^{\pi}\left[\cos\theta_0+\sin(\theta_0-\theta)\sin\theta\left(1-\cos(2\Omega t)\right)\right] e^{i(r-s)k}\;,
\\
\langle u^\dagger_r d_s \rangle
&=\frac{1}{4\pi}\int_{-\pi}^{\pi}\left[-\sin\theta_0+\sin(\theta_0-\theta)\left(\cos\theta\left(1-\cos(2\Omega t)\right)-i\sin(2\Omega t)\right)\right] e^{i(r-s)k}\;,
\\
\langle d^\dagger_r u_s \rangle
&=\frac{1}{4\pi}\int_{-\pi}^{\pi}\left[-\sin\theta_0+\sin(\theta_0-\theta)\left(\cos\theta\left(1-\cos(2\Omega t)\right)+i\sin(2\Omega t)\right)\right] e^{i(r-s)k}\;
\\
\langle d^\dagger_r d_s \rangle
&=\frac{\delta_{rs}}{2}+\frac{1}{4\pi}\int_{-\pi}^{\pi}\left[\cos\theta_0+\sin(\theta_0-\theta)\sin\theta\left(1-\cos(2\Omega t)\right)\right] e^{i(r-s)k}\;,
\end{split}
\end{align}
where we suppressed the momentum label of functions in the integrand for simplicity.
The spectrum of this $\langle \Psi^\dagger_r\Psi_s\rangle$, denoted by $\{\nu\}$ is related to (a double copy of) the normal modes of the reduced density matrix denoted by $\{\epsilon\}$ as $\nu_i=\pm\frac{1}{2}\tanh{\frac{\epsilon_i}{2}}$ and the EE is given by
\be
S_\ell(t)=\sum_{i=1}^{\ell}\left[\ln\left(1+e^{-\epsilon_i}\right)+\frac{\epsilon_i\,e^{-\epsilon_i}}{1+e^{-\epsilon_i}}\right]\;.
\ee

\section{Bosonic Model}
The bosonic model we have studied is the Lifshitz harmonic lattice model previously introduced in \cite{LHLM}. 
We briefly review this model here. The Lagrangian density of interest is \be
\mathcal{L}_b=\frac{1}{2}\left[\dot{\phi}^2-\left(\partial^z\phi\right)^2-m^{2z}\phi^2\right]\;.
\ee
The corresponding discrete Hamiltonian is given by
\be
H_b=\frac{1}{2}\sum_n\left[\frac{p^2_n}{\epsilon}+\epsilon\, m^{2z}\,q_n^2+\frac{1}{\epsilon^z}\left(\sum_{i=0}^z(-1)^{z+i}\;{}_zC_i\;q_{n+i-1}\right)\right]\;,
\ee
where ${}_zC_i=\frac{z!}{(z-i)!i!}$ and $z$ is any positive integer, the canonical commutation relation is $[q_r,p_s]=i\delta_{rs}$ and $\epsilon$ is the lattice spacing which we have fixed to unity. In the momentum basis this model is given by
\be
H_b=\int_{-\pi}^{\pi}dk\,\omega_k \left(a_{-k}^\dagger a_{-k}+a_k^\dagger a_k\right)\;,
\ee
where $\omega_k=\sqrt{m^{2z}+\left(2\sin k\right)^{2z}}$ and
$a^\dagger_k$ and $a_k$ are bosonic creation and annihilation operators.

To utilize the correlator method to find the time evolution of EE after quenching from $\omega_{0,k}$ to $\omega_k$, we plug the following correlators
\begin{align}
\begin{split}
X&\equiv\langle q_r q_s \rangle
=\frac{1}{2\pi}\int_{-\pi}^{\pi}
\frac{1}{\omega}\left[\frac{\omega}{\omega_0}\cos^2\omega t +\frac{\omega_0}{\omega}\sin^2\omega t\right] e^{i(r-s)k}\;,
\\
P&\equiv\langle p_r p_s \rangle
=\frac{1}{2\pi}\int_{-\pi}^{\pi}
\omega\left[\frac{\omega}{\omega_0}\sin^2\omega t +\frac{\omega_0}{\omega}\cos^2\omega t\right] e^{i(r-s)k}\;,
\\
R&\equiv\frac{1}{2}\langle \{q_r,p_s\} \rangle
=\frac{1}{2\pi}\int_{-\pi}^{\pi}
\left[\frac{\omega}{\omega_0}- \frac{\omega_0}{\omega}\right]\sin\omega t\;\cos\omega t \;e^{i(r-s)k}\;,
\end{split}
\end{align}
into 
\be
\Gamma=
\begin{pmatrix}
X & R \\ R^T & P
\end{pmatrix}
\;\;\;\;,\;\;\;\;
J=
\begin{pmatrix}
0 & \mathbf{1} \\ -\mathbf{1} & 0
\end{pmatrix}\;,
\ee
to read (a double copy of) the spectrum of the reduced density matrix denoted by $\{\epsilon\}$, via the spectrum of $iJ\cdot\Gamma$, denoted by $\{\lambda\}$ with
$\lambda_i=\pm\frac{1}{2}\coth{\frac{\epsilon_i}{2}}$. The EE is then given by
\be
S_\ell(t)=\sum_{i=1}^{\ell}\left[-\ln\left(1-e^{-\epsilon_i}\right)+\frac{\epsilon_i\,e^{-\epsilon_i}}{1-e^{-\epsilon_i}}\right]\;.
\ee


\begin{thebibliography}{99}
\bibitem{f0}
Our analysis for $z=1$ recovers the well-known results for CFT.
\bibitem{fLR}
No Lieb-Robinson bound has been proven so far for these theories. In theories that such a bound with an exponential tail exists, it is known that the leakage of information outside of effective light-cone is negligible \cite{LRBl}.
\end{thebibliography}
\end{document}